\documentstyle[12pt,epsfig]{article}

\begin{document}
\baselineskip=23pt

\vspace{1.2cm}

\begin{center}
{\Large \bf  Covariant Theory of Gravitation in the Spacetime with
Finsler Structure}

\bigskip

Xin-Bing Huang\footnote{huangxb@shnu.edu.cn}
\footnote{This paper is in memory of Prof. S.-S. Chern, a great mathematician I admire.}\\
{\em Shanghai United Center for Astrophysics (SUCA),} \\
{\em   Shanghai Normal University, No.100 Guilin Road, Shanghai
200234, China}
\end{center}

\bigskip
\bigskip
\bigskip

\centerline{\large Abstract} The theory of gravitation in the
spacetime with Finsler structure is constructed. It is shown that
the theory keeps general covariance. Such theory reduces to
Einstein's general relativity when the Finsler structure is
Riemannian. Therefore, this covariant theory of gravitation is an
elegant realization of Einstein's thoughts on gravitation in the
spacetime with Finsler structure.

\vspace{1.2cm}


\vspace{1.2cm}

\newpage

\section{Introduction}
In Newtonian mechanics, the spacetime is treated as a
three-dimensional Euclidean space and an independent
one-dimensional time. To solve the contradiction between the
Newtonian spacetime and the Maxwell's electromagnetic theory,
Einstein proposed the Minkowskian spacetime in his special
relativity. Furthermore, Einstein introduced four-dimensional
Riemannian spacetime in his gravitational field theory. After
that, large amount of astronomical observations have shown that
our spacetime is curved. Hence, the history of physics
demonstrates that the geometry of our spacetime should be decided
by the astronomical observations and the physical experiments.

At present our fundamental field theories are invariant under time
reversal. But our everyday feelings tell us that there do exist a
time arrow by which our world, our life and everything are
controlled. So the direction of time should play a role in our
theory on spacetime structure. The recent difficulties encountered
in solving dark energy problems also hint that the geometry of our
spacetime may be not Riemannian but its generalized case $--$
Finsler geometry.

In this paper, we try to construct a covariant field theory of
gravitation in the spacetime with the Finsler structure. We hope
that this theory will provide a new powerful platform for solving
the problems appeared in modern cosmology.

The paper is organized as follows: The Finsler structure in
four-dimensional spacetime is introduced, and by which the
fundamental tensor and the Cartan tensor are induced in Section 2.
In Section 3 we introduce the Chern connection, which is the
elegant mathematical tools in Finsler geometry. The curvatures of
Chern connection are discussed, and several kinds of Bianchi
identities are given in Section 4. We introduce a new tensor
${\cal Z}_{ijk}(x,y)$ in Section 5, by which a curvature-like
tensor ${\cal H}_{ijkl}(x,y)$ is introduced. The covariant field
equations of gravitation are acquired in this section also. In the
last section, we discuss the strong constraint given by the
conservation of the energy-momentum tensor. The vertical covariant
derivative of the tensor ${\cal Z}_{ijk}(x,y)$ is discussed also.

\section{Finsler Structure in four-dimensional Spacetime}

For the spacetime of physics we need four coordinates, the time
$t$ and three physical space coordinates $x^{1}$, $x^{2}$,
$x^{3}$. We put $t=x^{0}$, so that the four coordinates may be
written $x^{i}$, where the indice $i$ takes on the four values
$0$, $1$, $2$, $3$. Mathematically suppose that four-dimensional
spacetime can be treated as a $C^{\infty}$ manifold, denoted by
$M$. Denote by $T_{x}M$ the tangent space at $x\in M$, and by
$TM:={\cup}_{x\in M}~T_{x}M$ the tangent bundle of $M$. Each
element of $TM$ has the form $(x,y)$, where $x\in  M$ and $y\in
T_{x}M$. The natural projection $\pi:TM\to M$ is given by $\pi
(x,y):=x$. The dual space of $T_{x}M$ is $T^{*}_{x}M$, called the
cotangent space at $x$. The union $T^{*}M:={\cup}_{x\in
M}~T^{*}_{x}M$ is the cotangent bundle. Therefore,
$\{\frac{\partial}{\partial x^{i}}\}$ and $\{dx^{i}\}$ are,
respectively, the induced coordinate bases for $T_{x}M$ and
$T^{*}_{x}M$. The said $x^{i}$ give rise to local coordinates
$(x^{i},y^{i})$ on $TM$ through the mechanism\footnote{In this
paper, the rules that govern our index gymnastics are as follows:
\begin{enumerate}
\item Vector indices are up, and covector indices are down. \item
Any repeated pair of indices -- provided that one is up and the
other is down -- is automatically summed. \item The raising and
lowering of indices are carried out by the matrix $g_{ij}$ defined
by equation (\ref{f2}), and its matrix inverse $g^{ij}$.
\end{enumerate}}
\begin{equation}
\label{daor102} y=y^{i}\frac{\partial}{\partial x^{i}}~.
\end{equation}
The $y^{i}$ are fiberwise global. So functions $F(x,y)$ that are
defined on $TM$ can be locally expressed as
$F(x^{0},x^{1},x^{2},x^{3};y^{0},y^{1},y^{2},y^{3})$.

In Ref.\cite{bcs00} a Finsler structure of an n-dimensional
$C^{\infty}$ manifold is given from the mathematical point of
view. Similarly a Finsler Structure function of the
four-dimensional spacetime can be defined globally
\begin{equation}
\label{f1} F:~TM\to (-\infty, +\infty)
\end{equation}
with the following properties:
\begin{enumerate}
\item Regularity: $F(x,y)$ is $C^{\infty}$ on the entire slit
tangent bundle $TM_{o}:=TM {\backslash} \{0\}$.

\item Positive homogeneity: $F(x,\lambda y)=\lambda F(x,y)$ for
all $\lambda >0$.

\item The components of fundamental tensor:
\begin{equation}
\label{f2} g_{ij}(x,y)~:=~ \left[ \frac{1}{2}~F^{2}(x,y)
\right]_{y^{i}y^{j}}~,
\end{equation}
where
\begin{equation}
\label{f3}  \left[ \frac{1}{2}~F^{2}(x,y)
\right]_{y^{i}y^{j}}~:=~\frac{1}{2}~\frac{\partial^{2}
F^2(x,y)}{\partial y^{i}\partial y^{j}}~.
\end{equation}
In Finsler geometry, the $y^{i}$ (or $x^{i}$) appear in downstairs
indices usually denote the partial derivatives with respect to the
$y^{i}$ (or $x^{i}$). In the spacetime with Finsler structure, the
components of fundamental tensor can be decomposed by introducing
the tetrad matrix $e^{~a}_{i}(x,y)$ in local coordinates
\begin{equation}
\label{f4} g_{ij}(x,y)=e^{~a}_{i}(x,y)e^{~b}_{j}(x,y)\eta_{ab}~.
\end{equation}
we adopt the same sign conventions as that used by
Misner-Thorne-Wheeler's book\cite{mtw73}. The metric tensor of
local Minkowskian spacetime $\eta_{ab}$ is written as follows
\begin{equation}
\label{f0}
\eta_{00}=-1~,~~~~\eta_{11}=\eta_{22}=\eta_{33}=+1~,~~~~
\eta_{ab}=0~~~~{\rm for}~~~~a \neq b~,
\end{equation}
which is $not$ positive-definite.
\end{enumerate}
Given a Finsler structure $F(x,y)$ on the tangent bundle of
four-dimensional spacetime $M$, the pair $(M,F)$ can be called a
{\bf Finsler spacetime}.

In mathematical books on Finsler geometry, the Finsler structure
function $F$ satisfies $F\ge 0$ and the fundamental matrix
$g_{ij}(x,y)$ be positive-definite at every point of $TM_{o}$. But
in the case of four-dimensional spacetime, the tangent space is a
Minkowskian spacetime, therefore $F(x,y)$ may not satisfy
$F(x,y)\ge 0$ and $g_{ij}(x,y)$ is not positive-definite either.

Now we need the mathematical content of the pulled-back bundle
$\pi^{*}TM$ or its dual $\pi^{*}T^{*}$M, Here we don't explain how
to construct this bundle, for more detailed discussions on the
pulled-back bundle $\pi^{*}TM$ or its dual $\pi^{*}T^{*}$M, please
read Chapter 2 in Ref.\cite{bcs00}. For ease of local
computations, it is to our advantage to work on an affine
parameter space, where spacetime coordinates are readily
available. In this case, the preferred base manifold is the slit
tangent bundle $TM_{o}$. A good number of geometrical objects are
sections of the pulled-back bundle $\pi^{*}TM$ or its dual
$\pi^{*}T^{*}$M, or their tensor products. These sit over $TM_{o}$
and not $M$. Local coordinates $\{x^{i}\}$ on $M$ produce the
basis sections $\{\frac{\partial}{\partial x^{i}}\}$ and $\{
dx^{i}\}$, respectively, for $T_{x}M$ and $T^{*}_{x}M$. Now, over
each point $(x,y)$ on the manifold $TM_{o}$, the fiber of
$\pi^{*}TM$ is the vector space $T_{x}M$ while that of
$\pi^{*}T^{*}M$ is the covector space $T^{*}_{x}M$. Thus, the
$\frac{\partial}{\partial x^{i}}$ and $ dx^{i}$ give rise to
sections of the pulled-back bundles, in a rather simple-minded
way. In the pulled-back bundles these sections are defined locally
in $x$ and globally in $y$. This global nature in $y$ is automatic
because once $x$ is fixed, these sections do not change as we vary
$y$.

Hence a distinguished section ${\ell }$ of $\pi^{*}TM$ can be
defined by
\begin{equation}
\label{f5}
\ell=\ell_{(x,y)}:=\frac{y}{F}=\frac{y^{i}}{F(x,y)}\frac{\partial}{\partial
x^{i}}=: \ell^{i}\frac{\partial}{\partial x^{i}}~.
\end{equation}
Its natural dual is the Hilbert form $\omega$, which is a section
of $\pi^{*}T^{*}M$. We have
\begin{equation}
\label{f6}
\omega=\omega_{(x,y)}:=F_{y^{i}}(x,y)dx^{i}=F_{y^{i}}dx^{i}~.
\end{equation}
The definition (\ref{f5}) indicates that the components $\ell^{i}$
of the distinguished section $\ell$ satisfy
$\ell^{i}=\frac{y^{i}}{F}$. According to Euler's theorem in
Ref.\cite{bcs00}, it is obvious that
\begin{equation}
\label{f7} \ell_{i}:=g_{ij}(x,y)\ell^{j}=F_{y^{i}}(x,y)~.
\end{equation}
Thus the Hilbert form $\omega$ is expressible as
$\omega=\ell_{i}dx^{i}$. Both $\ell$ and $\omega$ are globally
defined on the manifold $TM_{o}$. The asserted duality means that
\begin{equation}
\label{f9} \omega(\ell)=\frac{y^{i}}{F}F_{y^{i}}=1~,
\end{equation}
which is a consequence of Euler's theorem too.

The pulled-back vector bundle $\pi^{*}TM$ admits a natural
Riemannian metric
\begin{equation}
\label{f11} g=g_{ij}dx^{i}\otimes dx^{j}~,
\end{equation}
where the components of $g$ is defined by equation (\ref{f2}),
obviously $g_{ij}=F F_{y^{i}y^{j}}+F_{y^{i}}F_{y^{j}}$ and
$g_{ij}=g_{ji}$. This is the fundamental tensor, which determines
the basic properties of the Finsler spacetime. It is a symmetric
section of $\pi^{*}T^{*}M\otimes \pi^{*}T^{*}M$. Likewise, another
important tensor in the Finsler spacetime is the Cartan tensor
\begin{equation}
\label{f13} A=A_{ijk}dx^{i}\otimes dx^{j}\otimes dx^{k}~,
\end{equation}
where the components is given by
\begin{equation}
\label{f14} A_{ijk}:=\frac{F}{2}\frac{\partial g_{ij}}{\partial
y^{k}}=\frac{F}{4} \left[ F^{2} \right]_{y^{i}y^{j}y^{k}}~,
\end{equation}
which is a totally symmetric section of $\pi^{*}T^{*}M\otimes
\pi^{*}T^{*}M\otimes \pi^{*}T^{*}M$. Mathematically the object
\begin{equation}
\label{f15} C_{ijk}:=\frac{A_{ijk}}{F}=\frac{1}{2}\frac{\partial
g_{ij}}{\partial y^{k}}
\end{equation}
is called the Cartan tensor in the geometric literature at large.
%

\section{Chern Connection and Covariant Derivatives}

The components $g_{ij}$ of the fundamental tensor defined in
equation (\ref{f2}) are functions on $TM_{o}$, and are invariant
under the positive rescaling in $y$. We use them to define the
formal Christoffel symbols of the second kind
\begin{equation}
\label{f20} \gamma^{i}_{~jk}:=\frac{g^{im}}{2}\left(\frac{\partial
g_{mk}}{\partial x^{j}}+\frac{\partial g_{jm}}{\partial
x^{k}}-\frac{\partial g_{jk}}{\partial x^{m}}\right)~,
\end{equation}
where $g^{ij}$ is the matrix inverse of $g_{ij}$, and also the
quantities
\begin{equation}
\label{f22} N^{i}_{~j}:=
\gamma^{i}_{~jk}y^{k}-C^{i}_{~jk}\gamma^{k}_{~rs}y^{r}y^{s}~.
\end{equation}
The above quantities $N^{i}_{~j}$ can be reexpressed as follows
\begin{equation}
\label{f25}\frac{ N^{i}_{~j}}{F}:=
\gamma^{i}_{~jk}\ell^{k}-A^{i}_{~jk}\gamma^{k}_{~rs}\ell^{r}\ell^{s}~,
\end{equation}
which is invariant under the positive rescaling $y\mapsto \lambda
y$.

Let
$[~x^{i}=x^{i}(\tilde{x}^{0},\tilde{x}^{1},\tilde{x}^{2},\tilde{x}^{3}),
i=0,1,2,3~]$ be a change of coordinates on spacetime.
Correspondingly, the chain rule gives
\begin{equation}
\label{f29}y^{i}=\frac{\partial x^{i}}{\partial
\tilde{x}^{j}}~\tilde{y}^{j}~.
\end{equation}
The tangent bundle of the manifold $TM$ has a local coordinate
basis that consists of the $\frac{\partial}{\partial x^{i}}$ and
the $\frac{\partial}{\partial y^{i}}$. However, under the
transformation on $TM$ induced by a coordinate change
$x\to\tilde{x}$, the vector $\frac{\partial}{\partial x^{i}}$
transforms in a complicated manner as follows:
\begin{equation}
\label{f32} \frac{\partial}{\partial\tilde{x}^{j}}=\frac{\partial
x^{i}}{\partial \tilde{x}^{j}}\frac{\partial}{\partial
x^{i}}+\frac{\partial^{2}x^{i}}{\partial
\tilde{x}^{j}\partial\tilde{x}^{k}}~\tilde{y}^{k}~\frac{\partial}{\partial
y^{i}}~.
\end{equation}
On the other hand, the vector $\frac{\partial}{\partial y^{i}}$
transforms simply
\begin{equation}
\label{f34} \frac{\partial}{\partial\tilde{y}^{j}}=\frac{\partial
x^{i}}{\partial \tilde{x}^{j}}\frac{\partial}{\partial y^{i}}~.
\end{equation}
The cotangent bundle of the manifold $T^{*}M$ has a local
coordinate basis $\{dx^{i},dy^{i}\}$. Here, under the said
coordinate change, the $dx^{i}$ behave simply
\begin{equation}
\label{f36} d\tilde{x}^{i}=\frac{\partial \tilde{x}^{i}}{\partial
x^{j}}dx^{j}
\end{equation}
while the $dy^{i}$ transform complicatedly
\begin{equation}
\label{f34} d\tilde{y}^{i}~=~\frac{\partial
\tilde{x}^{i}}{\partial
x^{j}}dy^{j}~+~\frac{\partial^{2}\tilde{x}^{i}}{\partial
x^{j}\partial x^{k}}~y^{k}~dx^{j}~.
\end{equation}
To avoid the complexity in the transformation equations
(\ref{f32}) and (\ref{f34}), furthermore, to obtain the coordinate
bases that transform as tensor under the said coordinate change,
Ref.\cite{bcs00} introduces two new symbols $\frac{\delta}{\delta
x^{i}}$ and $\frac{\delta y^{i}}{F}$ to replace
$\frac{\partial}{\partial x^{i}}$ and $dy^{i}$ respectively. The
$\frac{\delta}{\delta x^{i}}$ are defined by
\begin{equation}
\label{f36} \frac{\delta}{\delta
x^{i}}~:=~\frac{\partial}{\partial
x^{i}}~-~N^{j}_{~i}\frac{\partial}{\partial y^{j}}~,
\end{equation}
and the $\frac{\delta y^{i}}{F}$ are given by
\begin{equation}
\label{f38} \frac{\delta
y^{i}}{F}~:=~\frac{1}{F}~\left(dy^{i}~+~N^{i}_{~j}dx^{j}\right)~,
\end{equation}
which is invariant under positive rescaling in $y$. Note that
\begin{eqnarray}\nonumber
\label{f42} \frac{\delta}{\delta x^{i}} && \stackrel{natural~
dual}{----\rightarrow} ~~~~ dx^{i}~,
\\\nonumber
F\frac{\partial}{\partial y^{i}} && \stackrel{natural~
dual}{----\rightarrow} ~~~~\frac{\delta y^{i}}{F}~.
\end{eqnarray}
Therefore, we just introduce two new natural bases that are dual
to each other: 1, the bases $\{\frac{\delta }{\delta x^{i}},
F\frac{\partial}{\partial y^{i}}\}$ for the tangent bundle of
$TM_{o}$; 2, the bases $\{dx^{i}, \frac{\delta y^{i}}{F}\}$ for
the cotangent bundle of $TM_{o}$. The Ref.\cite{bcs00} indicates
that the horizontal subspace spanned by the $\frac{\delta}{\delta
x^{i}}$ is orthogonal to the vertical subspace spanned by the
$F\frac{\partial}{\partial y^{i}}$ .

The Chern connection is a linear connection that acts on the
pulled-back vector bundle $\pi^{*}TM$, sitting over the manifold
$TM_{o}$. It is $not$ a connection on the bundle $TM$ over $M$.
Nevertheless, it serves Finsler geometry in a manner that
parallels what the Levi-Civita connection (Christoffel symbol)
does for Riemannian geometry. Here we cite Chern theorem on Chern
connection in Ref.\cite{bcs00} in the following.

{\bf Chern Theorem:} Let (M,F) be a Finsler manifold. The
pulled-back bundle $\pi^{*}TM$ admits a unique linear connection,
called the Chern connection. Its connection forms are
characterized by the structural equations:
\begin{enumerate}
\item {\bf Torsion freeness:}
\begin{equation}
\label{f52} d(dx^{i})~-~dx^{j}\wedge
\omega^{~i}_{j}~=~-~dx^{j}\wedge \omega^{~i}_{j}~=~0~.
\end{equation}
\item {\bf Almost $g$-compatibility:}
\begin{equation}
\label{f53}
dg_{ij}~-~g_{kj}~\omega_{i}^{~k}~-~g_{ik}~\omega^{~k}_{j}~=~2~A_{ijm}~\frac{\delta
y^{m}}{F}~.
\end{equation}
\end{enumerate}
In fact, The torsion freeness is equivalent to the absence of
$dy^{k}$ terms in $\omega^{~i}_{j}$, that is to say
\begin{equation}
\label{f54} \omega^{~i}_{j}~=~\Gamma^{i}_{~jk}~dx^{k}~,
\end{equation}
together with the symmetry
\begin{equation}
\label{f55} \Gamma^{i}_{~jk}~=~\Gamma^{i}_{~kj}~.
\end{equation}
Furthermore, almost metric-compatibility implies that
\begin{equation}
\label{f56}
\Gamma^{i}_{~jk}~=~\gamma^{i}_{~jk}~-~g^{il}~\left(A_{ljm}~\frac{N^{m}_{~k}}{F}
~+~A_{klm}~\frac{N^{m}_{~j}}{F}~-~A_{jkm}~\frac{N^{m}_{~l}}{F}\right)~.
\end{equation}
Equivalently,
\begin{equation}
\label{f58} \Gamma^{i}_{~jk}~=~\frac{g^{il}}{2}~\left(\frac{\delta
g_{lk}}{\delta x^{j}}~+~\frac{\delta g_{jl}}{\delta
x^{k}}~-~\frac{\delta g_{jk}}{\delta x^{l}}\right)~,
\end{equation}
where the operators $\frac{\delta }{\delta x^{i}}$ have been
defined by equation (\ref{f36}).

Using the Chern connection, the covariant derivatives of the
tensors that are the sections of the pulled-back bundle
$\pi^{*}TM$ or its dual $\pi^{*}T^{*}M$, or their tensor product
can be calculated. For instance, let $T:=
T^{i}_{~j}\frac{\partial}{\partial x^{i}}\otimes dx^{j}$ be an
arbitrary smooth $(1,1)$-type tensor, which sits on the manifold
$TM_{o}$. Its covariant differential is
\begin{equation}
\label{f62} \nabla T ~:=~(\nabla
T)^{i}_{~j}~\frac{\partial}{\partial x^{i}}\otimes dx^{j}~,
\end{equation}
where $(\nabla T)^{i}_{~j}$ is
\begin{equation}
\label{f64} (\nabla
T)^{i}_{~j}~:=~dT^{i}_{~j}~-~T^{i}_{~k}~\omega^{~k}_{j}~+~T^{k}_{~j}~\omega^{~i}_{k}~.
\end{equation}
The components $(\nabla T)^{i}_{j}$ are $1$-forms on $TM_{o}$.
They are therefore be expanded in terms of the natural basis
$\{dx^{i}\}$ and $\{\frac{\delta y^{i}}{F}\}$, set
\begin{equation}
\label{f66} (\nabla T)^{i}_{~j}~=~T^{i}_{~j |
k}~dx^{k}~+~T^{i}_{~j;k}~\frac{\delta y^{k}}{F}~,
\end{equation}
where $T^{i}_{~j | k}$ is the horizontal covariant derivative of
$(\nabla T)^{i}_{~j}$ and $T^{i}_{~j;k}$ is the vertical covariant
derivative of $(\nabla T)^{i}_{~j}$ respectively. In order to
obtain formulas for the coefficients, we evaluate equation
(\ref{f66}) on each individual member of the dual basis
$\{\frac{\delta}{\delta x^{i}},~F\frac{\partial}{\partial
y^{i}}\}$. We also use the fact that the Chern connection forms
for the natural basis have no $\frac{\delta y^{k}}{F}$ terms, and
are given by equation (\ref{f54}). Therefore the results are
\begin{eqnarray}
\label{f70} && T^{i}_{~j | k}~=~\left(
\nabla_{\frac{\delta}{\delta x^{k}}}~T
\right)^{i}_{~j}~=~\frac{\delta T^{i}_{~j}}{\delta
x^{k}}~+~T^{l}_{~j}\Gamma^{i}_{~lk}~-~T^{i}_{~l}\Gamma^{l}_{~jk}~,
\\
&& T^{i}_{~j ; k}~=~\left( \nabla_{F\frac{\partial}{\partial
y^{k}}}~T \right)^{i}_{~j}~=~F\frac{\partial T^{i}_{~j}}{\partial
y^{k}}~.
\end{eqnarray}
The treatment for tensor fields of higher or lower rank is
similar. Here we list the covariant derivatives of several
important tensors. First Chern theorem says that the Chern
connection is almost $g$-compatible, namely
\begin{equation}
\label{f80} (\nabla
g)_{ij}~=~dg_{ij}~-~g_{kj}\omega^{~k}_{i}~-~g_{ik}\omega^{~k}_{j}~=~2~A_{ijl}~\frac{\delta
y^{l}}{F}~.
\end{equation}
This shows that the covariant derivatives of fundamental tensor
are
\begin{eqnarray}
\label{f81} && g_{~ij | l}~=~0~,
\\\label{f82}
&& g_{~ij ; l}~=~2~A_{ijl}~.
\end{eqnarray}
The obvious equations $(g^{ij}g_{jk})_{|l}~=~0$ and
$(g^{ij}g_{jk})_{;l}~=~0$ yield
\begin{equation}
\label{f83}
g^{ij}_{~~|l}~=~0~,~~~~~~g^{ij}_{~~;l}~=~-2A^{ij}_{~~l}~.
\end{equation}
Secondly, the covariant derivatives of the distinguished $\ell$
are
\begin{eqnarray}
\label{f85} && \ell^{i}_{~ | j}~=~0~,
\\\label{f86}
&& \ell^{i}_{~;j}~=~\delta^{i}_{~j}~-~\ell^{i}\ell_{j}~.
\end{eqnarray}
These, together with (\ref{f81}) and (\ref{f82}), can then be used
to deduce that
\begin{eqnarray} \label{f88} && \ell_{i |
j}~=~0~,
\\\label{f89}
&& \ell_{i;j}~=~g_{ij}~-~\ell_{i}\ell_{j}~.
\end{eqnarray}
Those covariant derivatives will be used in the process of
constructing the field equations of gravitation.

\section{Curvature and Bianchi Identities}
The curvature 2-forms of the Chern connection are
\begin{equation}
\label{f213}
\Omega^{~i}_{j}~:=~d\omega^{~i}_{j}-\omega^{~k}_{j}\wedge
\omega^{~i}_{k}~.
\end{equation}
Since the $\Omega^{~i}_{j}$ are 2-forms on the manifold $TM_{o}$,
Chern proved that they can be expanded as
\begin{equation}
\label{f216}
\Omega^{~i}_{j}~=~\frac{1}{2}~R^{~i}_{j~kl}~dx^{k}\wedge
dx^{l}~+~P^{~i}_{j~kl}~dx^{k}\wedge \frac{\delta y^{l}}{F}~.
\end{equation}
The objects $R$, $P$ are respectively the $hh-$, $hv-$curvature
tensors of the Chern connection. The wedge product $dx^{k}\wedge
dx^{l}$ in above equation demonstrates that
\begin{equation}
\label{f218} R^{~i}_{j~kl}~=~-~R^{~i}_{j~lk}~.
\end{equation}
The first Bianchi identities deduced from the torsion freeness of
the Chern connection uncovers a symmetry on $P^{~i}_{j~kl}$
\begin{equation}
\label{f222} P^{~i}_{j~kl}~=~P^{~i}_{k~jl}
\end{equation}
and the first Bianchi identity for $R^{~i}_{j~kl}$
\begin{equation}
\label{f225} R^{~i}_{j~kl}~+~R^{~i}_{k~lj}~+R^{~i}_{l~jk}~=~0~.
\end{equation}

In natural coordinates, formulas for $R^{~i}_{j~kl}$ and
$P^{~i}_{j~kl}$ are expressed in terms of the Chern connection
$\Gamma^{i}_{~jk}$ as follows
\begin{equation}
\label{f231} R^{~i}_{j~kl}~=~\frac{\delta \Gamma^{i}_{~jl}}{\delta
x^{k}}~-~\frac{\delta \Gamma^{i}_{~jk}}{\delta
x^{l}}~+~\Gamma^{i}_{~hk}\Gamma^{h}_{~jl}~-~\Gamma^{i}_{~hl}\Gamma^{h}_{~jk}~,
\end{equation}
and
\begin{equation}
\label{f236} P^{~i}_{j~kl}~=~-~F\frac{\partial
\Gamma^{i}_{~jk}}{\partial y^{l}}~.
\end{equation}
Note that equations (\ref{f55}) and (\ref{f236}) imply
(\ref{f222}).

The Chern connection is almost metric-compatible. Using this
property, some Bianchi identities are found. After exterior
differentiation on equation (\ref{f53}) and some manipulations, we
get
\begin{eqnarray}
\nonumber &~& \Omega_{ij}~+~\Omega_{ji}
\\\nonumber
&=& \frac{1}{2}~(R_{ijkl}~+~R_{jikl})~dx^{k}\wedge
dx^{l}~+~(P_{ijkl}~+~P_{jikl})~dx^{k}\wedge\frac{\delta y^{l}}{F}
\\\label{fxxx}
&=&-~(A_{iju}R^{u}_{~kl})~dx^{k}\wedge
dx^{l}~-~2(A_{iju}P^{u}_{~kl}~+~A_{ijl|k})~dx^{k}\wedge\frac{\delta
y^{l}}{F}
\\\nonumber
&~&+~2(A_{ijk;l}-A_{ijk}\ell_{l})~\frac{\delta
y^{k}}{F}\wedge\frac{\delta y^{l}}{F}~.
\end{eqnarray}
Here, we have introduced the abbreviations
\begin{eqnarray} \label{f226} &&
R^{i}_{~kl}~:=~\ell^{j}~R^{~i}_{j~kl}~.
\\\label{f227}
&&P^{i}_{~kl}~:=~\ell^{j}~P^{~i}_{j~kl}~.
\end{eqnarray}
There are three identities that one can obtain from above equation
(\ref{fxxx}). We carry them out systematically. The coefficients
of the $dx^{k}\wedge dx^{l}$ terms in (\ref{fxxx}) tell us that
\begin{equation}
\label{f336} R_{ijkl}~+~R_{jikl}~=~-~2~A_{iju}R^{u}_{~kl}~.
\end{equation}
The coefficients of the $dx^{k}\wedge\frac{\delta y^{l}}{F}$ terms
in (\ref{fxxx}) tell us that
\begin{equation}
\label{f338}
P_{ijkl}~+~P_{jikl}~=~-~2~A_{iju}P^{u}_{~kl}~-~2~A_{ijl|k}~.
\end{equation}
Apply (\ref{f338}) three times to the combination
\begin{equation}
\label{f342}
(P_{ijkl}~+~P_{jikl})~-~(P_{jkil}~+~P_{kjil})~+~(P_{kijl}~+~P_{ikjl})~.
\end{equation}
Through some operation, the result takes the form
\begin{equation}
\label{f342}
P_{jikl}~=~-~(A_{ijl|k}-A_{jkl|i}+A_{kil|j})~-~(A_{iju}P^{u}_{~kl}-A_{jku}P^{u}_{~il}+A_{kiu}P^{u}_{~jl})~.
\end{equation}
Contract this equation with $\ell^{j}$, respectively, adopt
(\ref{f85}) and the facts $P^{i}_{~jk}\ell^{k}=0$,
$\ell^{i}A_{ijk}=0$, we then reduces the contraction to the
important statement
\begin{equation}
\label{f356} P_{ikl}~:=~ \ell^{j}~P_{jikl}~=~- \dot{A}_{ikl}~.
\end{equation}
Here,
\begin{equation}
\label{f359} \dot{A}_{ikl}~:=~ A_{ikl|m}~\ell^{m}~.
\end{equation}
Then (\ref{f342}) and (\ref{f356}) together lead to the
constitutive relation for $P_{jikl}$
\begin{equation} \label{f366}
P_{jikl}~=~-(A_{ijl|k}-A_{jkl|i}+A_{kil|j})+(A_{ij}^{~~u}\dot{A}_{ukl}-A_{jk}^{~~u}\dot{A}_{uil}+A_{ki}^{~~u}\dot{A}_{ujl})~.
\end{equation}
Formula (\ref{f356}) can be used to reexpress equation
(\ref{f338}) as
\begin{equation} \label{f369}
A_{ijl|k}~=~A_{ij}^{~~u}\dot{A}_{ukl}~-~
\frac{1}{2}~\left(P_{ijkl}+P_{jikl}\right)~.
\end{equation}
Finally, the coefficients of the $\frac{\delta
y^{k}}{F}\wedge\frac{\delta y^{l}}{F}$ terms in (\ref{fxxx}) gives
\begin{equation} \label{f388}
A_{ijk;l}~-~A_{ijl;k}~=~A_{ijk}\ell_{l}~-~A_{ijl}\ell_{k}~.
\end{equation}
So we have discussed all formulas taken from (\ref{fxxx}).

Exterior differentiation of (\ref{f213}) gives the second Bianchi
identity
\begin{equation} \label{f388}
d\Omega^{~i}_{j}~-~\omega_{j}^{~k}\wedge\Omega_{k}^{~i}~+~\omega^{~i}_{k}\wedge\Omega^{~k}_{j}~=~0~.
\end{equation}
With (\ref{f54}) and (\ref{f216}), one can expand the above
equation as follows
\begin{eqnarray} \nonumber
0&=&\frac{1}{2}~\left(R^{~i}_{j~kl|t}~-~P^{~i}_{j~ku}R^{u}_{~lt}\right)~
dx^{k}\wedge dx^{l}\wedge dx^{t}
\\\nonumber
&+&\frac{1}{2}~\left(R^{~i}_{j~kl;t}~-~2~P^{~i}_{j~kt|l}~+~2~P^{~i}_{j~ku}\dot{A}^{u}_{~lt}\right)~dx^{k}\wedge
dx^{l}\wedge\frac{\delta y^{t}}{F}
\\\nonumber
&+&\left(
P^{~i}_{j~kl;t}~-~P^{~i}_{j~kl}\ell_{t}\right)~dx^{k}\wedge\frac{\delta
y^{l}}{F}\wedge\frac{\delta y^{t}}{F}~.
\end{eqnarray}
The above equation is equivalent to the following three
identities:
\begin{equation} \label{f422}
R^{~i}_{j~kl|t}~+~R^{~i}_{j~lt|k}~+~R^{~i}_{j~tk|l}~
=~P^{~i}_{j~ku}R^{u}_{~lt}~+~P^{~i}_{j~lu}R^{u}_{~tk}~+~P^{~i}_{j~tu}R^{u}_{~kl}~,
\end{equation}
\begin{equation} \label{f423}
R^{~i}_{j~kl;t}~=~P^{~i}_{j~kt|l}~-~P^{~i}_{j~lt|k}~-~
\left(P^{~i}_{j~ku}\dot{A}^{u}_{~lt}~-~P^{~i}_{j~lu}\dot{A}^{u}_{~kt}\right)~,
\end{equation}
\begin{equation} \label{f424}
P^{~i}_{j~kl;t}~-~P^{~i}_{j~kt;l}~=~P^{~i}_{j~kl}\ell_{t}~-~P^{~i}_{j~kt}\ell_{l}~.
\end{equation}
Making use of (\ref{f85}), Contracting (\ref{f422}) with
$\ell^{j}$, one can obtain
\begin{equation} \label{f586}
R^{i}_{~kl|t}~+~R^{i}_{~lt|k}~+~R^{i}_{~tk|l}~
=~-~\dot{A}^{i}_{~ku}R^{u}_{~lt}~-~\dot{A}^{i}_{~lu}R^{u}_{~tk}~-~\dot{A}^{i}_{~tu}R^{u}_{~kl}~.
\end{equation}

In general relativity, Einstein used the second Bianchi identity
in Riemannian geometry to deduce the important Einstein tensor,
the contraction of whose covariant derivative is vanished. As we
will show, the second Bianchi identity in Finsler geometry also
plays an important role in building the elegant field equations of
gravitation.

\section{The Field Equations of Gravitation in Finsler Spacetime}

We find that a tensor ${\cal Z}_{ijk}(x,y) dx^{i}\otimes
dx^{j}\otimes dx^{k}$ is necessary in constructing the covariant
field equations of gravitation. The tensor ${\cal Z}_{ijk}(x,y)
dx^{i}\otimes dx^{j}\otimes dx^{k}$ is described by
\begin{equation} \label{f566}
{\cal Z}_{ijl|k}~=~{\cal Z}_{iju}\dot{A}^{u}_{~kl}~-~P_{ijkl}~.
\end{equation}
By comparing (\ref{f369}) with (\ref{f566}), we obtain a simple
relationship
\begin{equation}
\label{f568} A_{ijk}~=~\frac{1}{2}~({\cal Z}_{ijk}~+~{\cal
Z}_{jik})~.
\end{equation}
Above equation shows that ${\cal Z}_{ijk}(x,y) dx^{i}\otimes
dx^{j}\otimes dx^{k}$ is closely related with the Cartan tensor
$A_{ijk}(x,y) dx^{i}\otimes dx^{j}\otimes dx^{k}$. From the
definition of Chern connection (\ref{f56}), one easily obtains
\begin{equation}
\label{add1} \Gamma_{ijk}~+~\Gamma_{jik}~=~\frac{\partial
g_{ij}}{\partial x^{k}}~-~2A_{ijl}~\frac{N^{l}_{~k}}{F}~.
\end{equation}
Combining (\ref{f566}), (\ref{f568}) and (\ref{add1}), we conclude
that
\begin{equation}
\label{add3} \Gamma_{ijk}~+~{\cal
Z}_{jil}~\frac{N^{l}_{~k}}{F}~=~\frac{1}{2}~\left(\frac{\partial
g_{ij}}{\partial x^{k}}~+~\frac{\partial g_{ik}}{\partial
x^{j}}~-~\frac{\partial g_{jk}}{\partial x^{i}}\right)~.
\end{equation}
Namely
\begin{equation}
\label{add4} \Gamma^{i}_{~jk}~+~{\cal
Z}^{~i}_{j~l}~\frac{N^{l}_{~k}}{F}~=~\gamma^{i}_{~jk}~.
\end{equation}
The formula (\ref{add4}) demonstrates that the tensor ${\cal
Z}^{~i}_{j~l}(x,y)$ vanishes when the Finsler structure is
Riemannian.

 We introduce a curvature-like tensor ${\cal H}_{ijkl}$ as
follows
\begin{equation}
\label{f576} {\cal H}_{ijkl}~:=~R_{ijkl}~+~{\cal
Z}_{iju}R^{u}_{~kl}~,
\end{equation}
which is a combination of $hh$-curvature tensor $R_{ijkl}$ and an
additional revised-curvature tensor ${\cal Z}_{iju}R^{u}_{~kl}$.
The definition (\ref{f576}) and formula (\ref{f336}) tell us that
\begin{equation}
\label{f578}{\cal H}_{ijkl}~=~-~{\cal H}_{jikl}~.
\end{equation}
The definition (\ref{f576}) and formula (\ref{f218}) tell us that
\begin{equation}
\label{f579}{\cal H}_{ijkl}~=~-~{\cal H}_{ijlk}~.
\end{equation}

In the curvature-like tensor ${\cal H}_{ijkl}$, the second Bianchi
identity on $R_{ijkl}$ has been given, namely equation
(\ref{f422}). Now we try to find a similar identity on ${\cal
Z}_{j~u}^{~i}R^{u}_{~kl}$. First
\begin{equation}
\label{f578} \left({\cal
Z}_{j~u}^{~i}R^{u}_{~kl}\right)_{|t}~=~{\cal
Z}_{j~u|t}^{~i}R^{u}_{~kl}~+~{\cal Z}_{j~u}^{~i}R^{u}_{~kl|t}~.
\end{equation}
Apply (\ref{f578}) three times to the following combination
\begin{equation}
\label{f579} \left({\cal
Z}_{j~u}^{~i}R^{u}_{~kl}\right)_{|t}~+~\left({\cal
Z}_{j~u}^{~i}R^{u}_{~lt}\right)_{|k}~+~\left({\cal
Z}_{j~u}^{~i}R^{u}_{~tk}\right)_{|l}~.
\end{equation}
With the help of (\ref{f586}), (\ref{f566}) and (\ref{f578}), we
can get the result of expression (\ref{f579}) as follows
\begin{eqnarray} \nonumber
&~&\left({\cal Z}_{j~u}^{~i}R^{u}_{~kl}\right)_{|t}~+~\left({\cal
Z}_{j~u}^{~i}R^{u}_{~lt}\right)_{|k}~+~\left({\cal
Z}_{j~u}^{~i}R^{u}_{~tk}\right)_{|l}
\\\label{f600}
&=&~-~P^{~i}_{j~ku}R^{u}_{~lt}~-~P^{~i}_{j~lu}R^{u}_{~tk}~-~P^{~i}_{j~tu}R^{u}_{~kl}~.
\end{eqnarray}
Therefore, combining (\ref{f422}) with (\ref{f600}), we have an
important identity
\begin{equation} \label{604}
{\cal H}^{~i}_{j~kl|t}~+~{\cal H}^{~i}_{j~lt|k}~+~{\cal
H}^{~i}_{j~tk|l}~ =~0~.
\end{equation}
We shall be particularly concerned with the contracted form of
(\ref{604}). Recalling that the horizontal covariant derivatives
of $g^{jl}$ vanish, we find on contraction of $j$ with $l$ that
\begin{equation} \label{606}
{\cal H}^{i}_{~k|t}~-~{\cal H}^{i}_{~t|k}~+~{\cal
H}^{ji}_{~~tk|j}~ =~0~,
\end{equation}
where we have adopted the definition\footnote{We strongly feel
that ${\cal H}_{kj}$ satisfy ${\cal H}_{kj}={\cal H}_{jk}$, but
till now we cannot prove it. But we believe that geometers can
find it. If not, we have to redefine ${\cal H}_{ijkl}$ as follows
\begin{equation}
\label{fyy} {\cal
H}_{ijkl}~:=\frac{1}{2}~\left(R_{ijkl}~+~R_{klij}~+~{\cal
Z}_{iju}R^{u}_{~kl}~+~{\cal Z}_{klu}R^{u}_{~ij}\right)~.
\end{equation}
This re-definition will not change our field equations. But the
theory will be slightly ugly.}
\begin{equation}
\label{f589} {\cal H}_{kj}~:=~g^{il}{\cal H}_{ikjl}~=~{\cal
H}^{i}_{~kji}~.
\end{equation}
Contracting equation (\ref{606}) again gives
\begin{equation}
\nonumber
{\cal H}_{|t}~-~{\cal H}^{i}_{~t|i}~-~{\cal H}^{j}_{~t|j}~ =~0~,
\end{equation}
or
\begin{equation} \label{608}
\left({\cal H}^{j}_{~t}~-~\frac{1}{2}~\delta^{j}_{~t}~{\cal
H}\right)_{|j} =~0~,
\end{equation}
where the scalar ${\cal H}$ is
\begin{equation}
\label{f614} {\cal H}~:=~g^{ij}{\cal H}_{ij}~=~{\cal H}^{i}_{~i}~.
\end{equation}
An equivalent but more familiar form for (\ref{608}) is
\begin{equation} \label{609}
\left({\cal H}^{jt}~-~\frac{1}{2}~g^{jt}~{\cal H}\right)_{|j}
=~0~.
\end{equation}

In general relativity, the energy-momentum tensor $T_{ij}$ is
conserved to make sure the conservation of energy and
momentum~\cite{wei72}. Similarly, we introduce the energy-momentum
tensor ${\cal T}_{ij}(x,y) dx^{i}\otimes dx^{j}$ in the manifold
$\pi^{*}TM$. The covariant derivatives of the energy-momentum
tensor are
\begin{equation}
\label{fty1} (\nabla {\cal T})^{ij}~=~{\cal T}^{ij}_{~~ |
k}~dx^{k}~+~{\cal T}^{ij}_{~~;k}~\frac{\delta y^{k}}{F}~,
\end{equation}
where ${\cal T}^{ij}_{~~ | k}$ are the horizontal covariant
derivatives of ${\cal T}_{ij}(x,y)$ and ${\cal T}^{ij}_{~~;k}$ are
the vertical covariant derivatives of ${\cal T}_{ij}(x,y)$. When
the Finsler structure is Riemannian, ${\cal T}^{ij}_{~~;k}$
vanish, and the horizontal covariant derivatives ${\cal
T}^{ij}_{~~ | k}$ become the covariant derivatives of the
energy-momentum tensor. Therefore, from the physical point of
view, the energy and momentum conservation can be kept when the
horizontal covariant derivatives ${\cal T}^{ij}_{~~ | k}$ of the
energy-momentum tensor satisfy
\begin{equation}
\label{fty3} {\cal T}^{ij}_{~~~ | i}~=~0~.
\end{equation}
Frankly speaking, we still do not understand the physical meaning
of the vertical covariant derivatives of the energy-momentum
tensor.

With the energy-momentum tensor ${\cal T}_{ij}(x,y)$ sitting over
the Finsler spacetime, when its horizontal derivatives satisfy
(\ref{fty3}), we propose the covariant field equations of
gravitation as follows
\begin{equation}
\label{f611} {\cal H}^{ij}~-~\frac{1}{2}~g^{ij}~{\cal H}~=~8\pi~ G
~{\cal T}^{ij}~,
\end{equation}
where $G$ is the Newtonian constant. Obviously, (\ref{609}) makes
sure that the energy-momentum tensor in (\ref{f611}) satisfies
(\ref{fty3}). Therefore, the field equations (\ref{f611}) do
reserve the energy and momentum conservation.

Obviously, for the empty spacetime, the energy-momentum tensor in
(\ref{f611}) disappears, the field equations of gravitation reduce
to
\begin{equation}
\label{fempty} {\cal H}~=~0~.
\end{equation}

When the Finsler structure is Riemannian, the tensor ${\cal
Z}_{iju}R^{u}_{~kl}$ vanishes and the curvature-like tensor ${\cal
H}_{ijkl}$ becomes the Riemannian curvature tensor of Riemannian
spacetime, our field equations becomes Einstein's field equations
exactly. Therefore, our covariant field equations of gravitation
in the Finsler spacetime are the natural result of Einstein's
thoughts on gravitation.

\section{More Discussions}
The equation (\ref{fty1}) demonstrates that the energy-momentum
conservation in Finsler spacetime can be divided into two kinds.
First, the weak energy-momentum conservation, used in last
section. In this kind of energy-momentum conservation, we require
that ${\cal T}^{ij}_{~~~ | i}=0$, the field equations of
gravitation has been discussed. Secondly, the strong
energy-momentum conservation, we call, that is, both ${\cal
T}^{ij}_{~~ | i}$ and ${\cal T}^{ij}_{~~;i}$ vanish. The
constraint condition
\begin{equation}
\label{fvanish} {\cal T}^{ij}_{~~;i}~=~0
\end{equation}
sets a serious constraint on the vertical covariant derivatives of
tensor ${\cal Z}_{ijk}$. If conservation (\ref{fvanish}) is
needed, the field equations (\ref{f611}) tell us that
\begin{equation} \label{609**}
\left({\cal H}^{ij}~-~\frac{1}{2}~g^{ij}~{\cal H}\right)_{;i}
=~0~.
\end{equation}
But we cannot draw a concise equation on ${\cal Z}_{ijk;l}$ from
(\ref{609**}). We find that the constraint equation of ${\cal
Z}_{ijk;l}$ given by (\ref{609**}) is so complicated that no
valuable information can be obtained. Maybe, further studies will
demonstrate that (\ref{f566}) and (\ref{609**}) are exclusive,
then the strong energy-momentum conservation cannot be required in
the spacetime with the Finsler structure.

In a conclusion, we use the Chern connection and the curvatures
given by the Chern connection to set up the field equations of
gravitation in the Finsler spacetime. After introducing the tensor
${\cal Z}_{ijk}$ and its related curvature-like tensor ${\cal
H}^{~i}_{j~kl}$, we obtain the field equations of gravitation,
formally like Einstein's field equations. We show that our field
equations exactly reduces to Einstein's field equations when the
Finsler structure is Riemannian.



 {\bf Acknowledgement:}  I would like to thank Prof. Zhongmin Shen, Prof. Shenglin Cao
 and Dr. Benling Li for their help and enlightening discussions.

\end{document}